\documentclass[aps,prb,twocolumn,superscriptaddress,showpacs]{revtex4}
\usepackage{float}
\usepackage{graphicx}
\usepackage{dcolumn}
\usepackage{bm}
\usepackage{color}
\usepackage{float}
\usepackage{amsmath}
\usepackage{amssymb}

\newcommand{\eph}{{\it e-ph}}
\newcommand{\ee}{{\it e-e}}
\allowdisplaybreaks

\begin{document}
\title{Phase competition in a one-dimensional three-orbital Hubbard-Holstein model}
\author{Shaozhi Li}
\affiliation{Department of Physics and Astronomy, The University of Tennessee, Knoxville, Tennessee 37996, USA}
\author{Yanfei Tang}
\affiliation{Department of Physics, Virginia Tech, Blacksburg, Virginia 24061, USA}
\author{Thomas A. Maier}
\affiliation{Center for Nanophase Materials Sciences, Oak Ridge National Laboratory, Oak Ridge, Tennessee 37831, USA}
\affiliation{Computational Sciences and Engineering Division, Oak Ridge National Laboratory, Oak Ridge, Tennessee 37831, USA}
\author{Steven Johnston}
\affiliation{Department of Physics and Astronomy, The University of Tennessee, Knoxville, Tennessee 37996, USA}
\affiliation{Joint Institute of Advanced Materials at The University of Tennessee, Knoxville, Tennessee 37996, USA}

\begin{abstract}  
We study the interplay between the electron-phonon ({\eph}) and on-site electron-electron ({\ee}) interactions in a three-orbital Hubbard-Holstein model on an extended one-dimensional lattice using determinant quantum Monte
Carlo. For weak {\ee} and {\eph} interactions, we observe a competition between an orbital-selective Mott phase (OSMP) and a (multicomponent) charge-density-wave (CDW) insulating phase, with an intermediate metallic phase located between them. For large {\ee} and {\eph} couplings, the OSMP and CDW phases persist, while the metallic phase develops short-range orbital correlations and becomes insulating when both the {\ee} and {\eph} interactions are large but comparable. Many of our conclusions are in line with those drawn from a prior dynamical mean field theory study of the two-orbital Hubbard-Holstein model [Phys. Rev. B {\bf 95}, 12112(R) (2017)] in infinite dimension, suggesting that the competition between the {\eph} and {\ee} interactions in multiorbital Hubbard-Holstein models leads to rich physics, regardless of the dimension of the system. 
\end{abstract}

\date{\today}
\maketitle

\section{Introduction}  
The multiorbital iron-based superconductors (FeSCs) remain as a  significant research problem in the condensed matter physics
[\onlinecite{JohnstonAP, Stewart, DaiNat}]. While most of the FeSCs host
quasi-two-dimensional FeAs or FeSe layers, a quasi-one-dimensional (1D) structure has been discovered recently in $\mathrm{BaFe}_2\mathrm{S}_3$
[\onlinecite{Takahashinature}], which has spurred interest in quasi-1D multiorbital models [\onlinecite{NiravPRB2016, KaushalPRB, RinconPRL, LiuPRE}]. $\mathrm{BaFe}_2\mathrm{S}_3$  has the geometry of a two-leg ladder, and its
low-temperature magnetic structure is ferromagnetic (FM) in the ladder direction and antiferromagnetic (AFM) in the leg direction
[\onlinecite{Takahashinature, NiravPRB2016}]. Based on its magnetic properties,
and the partially filled Fe orbitals at the Fermi level
[\onlinecite{NiravPRB2016, ZhangPRB2017}], it is natural to believe that the
electron-electron ({\ee}) interactions play a critical role in determining the
characteristics of $\mathrm{BaFe}_2\mathrm{S}_3$ and other FeSCs
[\onlinecite{Dagotto_mod}]. 

To date, theoretical studies of correlated multiorbital models have revealed
many new phenomena such as a Hund's metal [\onlinecite{HauleNJP,
Georges, Fanfarillo_PRB2015}] and the orbital-selective Mott phase (OSMP)
[\onlinecite{Anisimov, Koga, deMediciPRL2014, Yi_PRL2013, ShaozhiPRB2016}].
These concepts are central to our understanding of the FeSCs.  At the same time, however, there is some evidence that the {\eph}
interaction can also be important in these materials. For example, {\eph}
interactions have been proposed as pairing mediators in  FeSe monolayers grown
on oxide substrates [\onlinecite{Liu_nature2012,
Lee_nature2014,DHLee2015,Xiang2016,Rademaker2016,KenPRL2017}], although the idea is under debate
[\onlinecite{SongPreprint,JandkePreprint,LiPreprint}]. Moreover, there is
experimental evidence that the low-energy electronic properties of the FeSC are
modified by the {\eph} interaction, as inferred from Raman
[\onlinecite{RahlenbeckPRB}] and infrared spectroscopy
[\onlinecite{XuPRB2015}], optical conductivity measurements
[\onlinecite{DongPRB2010}], and time-resolved photoemission
[\onlinecite{GerberScience}]. On a theoretical front, early {\it ab initio} calculations found that the {\eph} coupling strength in the FeSCs is minimal,
with total dimensionless couplings $\lambda \sim 0.2$, and insufficient to
establish the observed superconducting transition temperatures $T_c$
[\onlinecite{BoeriPRL2008}]. Several follow-up studies have examined the influence
of electronic and magnetic correlations and orbital degrees of freedom on the {\eph}
interaction in multiorbital systems [\onlinecite{SaitoPRB2010, YndurainPRB2009,
LiPreprint,Shaozhiarxiv,Coh, GerberScience}]. Each has found non-trivial effects and enhanced 
couplings, indicating that
the role of the {\eph} interaction may be more subtle than initially expected.
Studying {\eph} interactions in multiorbital materials is, therefore, an open
problem.  This issue is also relevant to the quasi-1D organic superconductors
of the Bechgaard and Fabre salt families [\onlinecite{Dupuis_ltp}]. 

A simplified class of models describing the combined {\eph} and
Coulomb interactions are Hubbard-Holstein (HH) models.  In the single-band case, the competition between the local Coulomb repulsion and the attractive
{\eph} interaction leads to a phase transition between antiferromagnetic and
charge-density-wave (CDW) order at half filling [\onlinecite{BauerPRB, NowadnickPRL,
MurakamoPRB2013,PradhanPRB2015,ClayPRL2005,HardikarPRB2007,KollerPRB2004}].
This phase transition has been studied extensively in different dimensions
[\onlinecite{BauerPRB, ClayPRL2005, NowadnickPRL}], as a function of doping
[\onlinecite{ClayPRL2005,HardikarPRB2007,MendlPRB2017}], and using many different methods
[\onlinecite{BauerPRB, NowadnickPRL}]. For example, an early dynamical mean
field theory (DMFT) study on the Bethe lattice found this phase transition is
continuous for small {\eph} couplings and discontinuous for large couplings
[\onlinecite{BauerPRB, JeohPRB}]. This conclusion differs from that of
determinant quantum Monte Carlo (DQMC) calculations on a two-dimensional
lattice, which found that the phase transition remains continuous for large couplings at high temperatures [\onlinecite{NowadnickPRL, JohnstonPRB2013}].
The phase transition in the one-dimensional HH model is also complicated. No
Mott phase transition occurs in the absence of the {\eph} interaction, and the
Mott insulating phase exists for any onsite Coulomb repulsion $U>0$
[\onlinecite{Lieb_PRL1968}]. Conversely, in the absence of {\ee} interactions,
the critical value of the {\eph} coupling strength exists for the formation of
a CDW phase due to quantum fluctuations of the phonon
field [\onlinecite{HirschPRL, HirschPRB, Bursill_PRL1998, Jeckelmann_PRB1999}].
Retardation effects are also more prominent in one dimension at strong couplings
[\onlinecite{HirschPRL, Fradkin_PRB1983}] when compared with the infinite-dimensional case [\onlinecite{BauerPRB}]. Thus, while a variety of techniques find a robust phase transition between the CDW and the Mott phases in all dimensions, the details of the transition can differ in important ways. It is
therefore important to study HH models in different dimensions and to use different approaches to obtain a complete physical picture.

Another critical factor in the single-band HH model is the degree of electron
doping. For example, at half filling, the CDW and antiferromagnetic Mott phases
are separated by a metallic phase at weak {\eph} coupling on the Bethe lattice
[\onlinecite{JeohPRB}]. But at small hole doping, phase separation
of the CDW and the metallic phases [\onlinecite{CaponePRL2004}] occurs.
In two-dimensions and at a quarter filling, a charge-ordered antiferromagnetic
phase appears for strong Coulomb repulsion instead of the usual ${\bf Q} =
(\pi/a,\pi/a)$ antiferromagnetic phase [\onlinecite{KumarPRB}]. For quarter
filling in 1D, and at weak {\eph} coupling, the antiferromagnetic phase
separates into an antiferromagnetic phase and a correlated singlet phase
[\onlinecite{RejaPRB}]. These results indicate that many new states can arise
in the doped single-band HH model. It is then natural to wonder how orbital
degrees of freedom enter into this problem. 

To date, multiorbital extensions of the HH model have received comparably 
less attention [\onlinecite{KontaniPRL,YamadaJPSJ,CaponeScience,CaponePRL2001,HanPRL,Shaozhiarxiv}]. In Ref.
[\onlinecite{Shaozhiarxiv}], some of the current authors examined orbital selective behaviors in a 
degenerate two-orbital HH model (with inequivalent bandwidths) at half filling
using single-site DMFT [\onlinecite{Shaozhiarxiv}]. We found that the combined {\eph} and onsite {\ee} interactions resulted in many competing phases including metallic, Mott, and CDW insulating phases, an OSMP, and a lattice-driven orbital-selective Peierls insulator (OSPI) analogue to the OSMP.  

Our prior study was carried out in infinite dimensions, where DMFT is exact; however, given the dependence on dimensionality found for the single-orbital HH 
model, it is essential to study the problem in other dimensions, as a function of doping, 
and with different techniques. Motivated by this, here we present a complementary study
of the three-orbital HH model defined on an extended 1D chain at an average electron filling $\langle\hat{n}\rangle = 4$. We study the model using DQMC, which is a
nonperturbative auxiliary-field technique capable of handling both the
{\ee} and {\eph} interactions on equal footing. 

\begin{figure}
 \includegraphics[width=0.78\columnwidth]{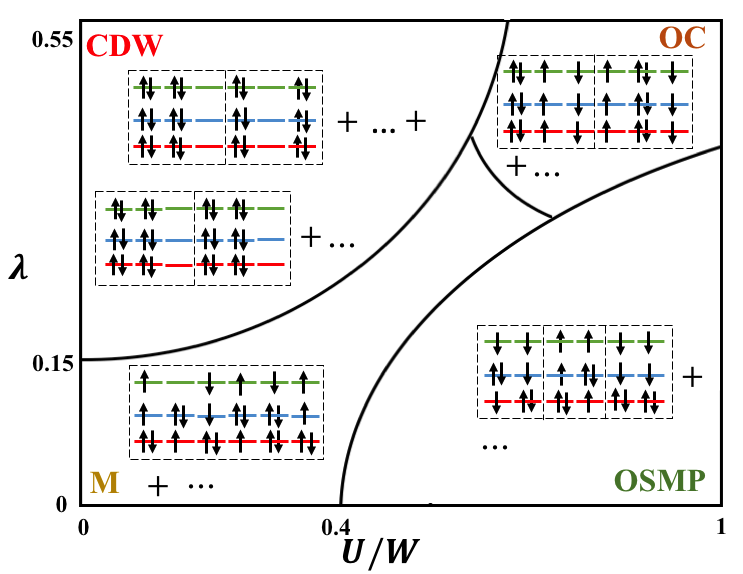}
 \caption{\label{Fig:Phase_sketch} (color online) 
 A sketch of the $\lambda$-$U$ phase diagram for our model as inferred from our 
 DQMC calculations. Four distinct regions are found, which include states with 
 metallic (M) characteristics, an orbital-selective Mott region (OSMP), a charge-density-wave 
 (CDW) order, and a region with strong orbital correlations (OC) and insulating characteristics. 
 The level diagrams sketch the dominant electronic configurations in each region. 
 }
\end{figure}

Previous studies of this three-orbital 1D Hubbard model found that it  
exhibits a rich variety of phases including block antiferromagnetism,
antiferromagnetism, and several types of OSMPs [\onlinecite{RinconPRL, LiuPRE,
ShaozhiPRB2016}]. As such, the model provides an excellent starting point for studying the impact of the {\eph} coupling on such physics. 
Our results show that the {\eph} interaction further modifies the boundaries of
these phases by introducing additional competition
with the CDW tendencies. By working on an extended
lattice, we can study momentum resolved quantities. Our main results are
summarized in Fig. \ref{Fig:Phase_sketch}, which provides a schematic of the
phase diagram of the model. 
(A more precise diagram with quantitative boundaries is provided at the end of our analysis, see Fig. \ref{Fig:fig10}.) 
We find four distinct regions in the $\lambda$-$U$ parameter space, 
where $\lambda$ and $U$ parameterize the {\eph} and {\ee} interactions, 
respectively. 
For small $\lambda$ and large $U$ we find an OSMP phase consistent with a block antiferromagnetic OSMP reported in prior work [\onlinecite{RinconPRL,LiuPRE}]. 
For large $\lambda$ and small $U$ we find an insulating CDW phase with mixed $q$ ordering. These two phases are separated 
by a metallic (M) phase at weak $\lambda$ and $U/W$, which persists into the region where $U$ and $\lambda$ are comparable to one another; 
however, if these same interactions are made sufficiently large but still 
comparable, the metallic state 
develops short-range orbital correlations (OC), resulting in insulating behavior. 
Our results show that metallicity can be lost in the presence of large {\ee} and {\eph} interactions due to the formation of short-range correlations. We also demonstrate that the 
addition of {\eph} interactions to multiorbital Hubbard models can strongly
influence the phases of the model and that this occurs irrespective of the
dimension, filling, or mechanism underlying other orbital-selective 
behaviors. 

\section{Methods}\label{Sec:Methods}
\subsection{Model Hamiltonian} 
Our starting point is a simplified one-dimensional three-orbital Hubbard model,
first introduced in Ref. [\onlinecite{RinconPRL}]. 
We then add a Holstein-type interaction, 
where the atomic displacement is coupled to the electron density on each
orbital. The full Hamiltonian is $H=H_0+H_{e-e}+H_\mathrm{lat}+H_{e-ph}$,
where
\begin{eqnarray}
H_0=-\sum_{\substack{\langle i, j \rangle \\ \sigma,\gamma,\gamma^\prime}}t^{\phantom\dagger}_{\gamma\gamma^\prime}c^{\dagger}_{i,\gamma,\sigma}c^{\phantom\dagger}_{j,\gamma^\prime,\sigma}+\sum_{i,\sigma,\gamma} (\Delta_\gamma-\mu) \hat{n}_{i,\gamma,\sigma}
\end{eqnarray}
are the non-interacting electronic terms,
\begin{eqnarray}
H_\mathrm{lat}=\sum_i\left[ \frac{\hat{P}_i^2}{2M}+\frac{M\Omega^2}{2}\hat{X}_i^2\right]=\Omega\sum_{i} \left( b^{\dagger}_ib^{\phantom\dagger}_i+\frac{1}{2}\right)
\end{eqnarray}
are the non-interacting lattice terms, 
\begin{eqnarray}\label{Eq:Hee}
H_{e-e}&=&U\sum_{i,\gamma}\hat{n}_{i,\gamma,\uparrow} \hat{n}_{i,\gamma,\downarrow}+\left( U^\prime-\frac{J}{2}\right) \sum_{\substack{i,\sigma,\sigma^\prime \\ \gamma<\gamma^\prime}}\hat{n}_{i,\gamma,\sigma}\hat{n}_{i,\gamma^\prime,\sigma^\prime}\nonumber\\
&&+J\sum_{i,\gamma<\gamma^\prime}S_{i,\gamma}^{z}S_{i,\gamma^\prime}^z
\end{eqnarray}
are the on-site Hubbard and Hund's interaction terms, and 
\begin{eqnarray}
H_{e-ph}=\alpha\sum_{i,\gamma,\sigma}\hat{X}_i\hat{n}_{i,\gamma,\sigma}=g\sum_{i,\gamma,\sigma} \left (b^{\dagger}_i + b^{\phantom\dagger}_i \right) \hat{n}_{i,\gamma,\sigma}
\end{eqnarray}
are the {\eph} coupling terms. Here, $\langle \dots \rangle$ denotes a
sum over nearest neighbors; $c^{\dagger}_{i,\gamma,\sigma}$ ($
c^{\phantom\dagger}_{i,\gamma,\sigma}$) creates (annihilates) a spin $\sigma$
electron in orbital $\gamma=$ 1, 2, 3 on site $i$; $b^{\dagger}_i$ ($
b^{\phantom\dagger}_i$) creates (annihilates) a phonon on lattice site $i$; $S_{i,\gamma}^{z}$
is the z-component of the spin operator ${\bf S}_{i,\gamma}$; 
$\hat{n}_{i,\gamma,\sigma}=c^\dagger_{i,\gamma,\sigma}c^{\phantom\dagger}_{i,\gamma,\sigma}$
is the number operator; and $\hat{X}_i$ and $\hat{P}_i$ are the lattice position and momentum
operators, respectively.  
The parameters $\Delta_\gamma$ are the on-site energies for each orbital; $t_{\gamma,\gamma^\prime}$ 
are the intra- and interorbital hopping integrals; $U$ and $U^\prime$ are the intra- and interorbital
Hubbard interactions, respectively, and $J$ is the Hund's coupling. The 
parameter $g=\frac{\alpha}{\sqrt{ 2M\Omega}}$ is the strength of the
{\eph} coupling and $\Omega$ is the phonon energy. Finally, $\mu$ is the chemical
potential, which fixes the average particle number. 

In Eq. (\ref{Eq:Hee}) we have neglected the pair-hopping
and spin-flip terms of the Hund's interaction, as was done in Ref. [\onlinecite{ShaozhiPRB2016}] for the same model without {\eph} interactions. These terms
introduce a significant Fermion sign problem [\onlinecite{LohPRB}] for our DQMC calculations and are therefore neglected to make the
problem tractable. Prior work [\onlinecite{LiuPRE}] has shown that these terms only change the location of the various phase boundaries for the model considered here in the absence of the {\eph} interaction. 
We, therefore, proceed assuming that this will also hold true
once the phonons are included in our calculations. 

\subsection{Model parameters}
Throughout this work, we choose $U'=U-2J$, as is standard for enforcing
rotational symmetry [\onlinecite{CastellaniPRL}], although we
have neglected the pair-hopping and spin-flip terms in Eq. (\ref{Eq:Hee}).  
We further vary $U$ while holding $J=U/4$ fixed. This choice produces a robust OSMP
[\onlinecite{RinconPRL, LiuPRE, ShaozhiPRB2016}] in the absence of the {\eph}
interaction and is appropriate for the FeSCs. 
We work at a fixed filling $\langle
\hat{n} \rangle =4$, which is typical for three-orbital Hubbard models used to
describe the 2D FeSCs [\onlinecite{DaghoferPRB2010}]. We expect that the same filling is needed to
describe the quasi-one-dimensional system $\mathrm{BaFe}_2\mathrm{S}_3$. 
This choice of filling also allows us to make direct comparisons to previous studies in the absence of the {\eph} interaction [\onlinecite{RinconPRL, LiuPRE, ShaozhiPRB2016}]. 
In this spirit, we also set $t_{11}=t_{22}=-0.5$ eV, $t_{33}=-0.15$ eV,
$t_{13}=t_{23}=0.1$ eV, $t_{12}=0$ eV, $\Delta_1=-0.1$ eV, $\Delta_2=0$ eV, $\Delta_3=0.8$ eV, and $\Omega=0.5$ eV, 
again following 
Refs. [\onlinecite{ShaozhiPRB2016}] and  [\onlinecite{RinconPRL}]. The total
bandwidth of the non-interacting model is $W=2.45$ eV, which serves as the unit
energy. The dimensionless {\eph} coupling
constant is defined as $\lambda=\alpha^2/(M\Omega^2W)$. (Note that since this
is a multi-band system, different choices of bandwidths are possible. Here, we
select the total bandwidth, as was done in Ref. [\onlinecite{Shaozhiarxiv}].) 
Finally, we set $a=M=1$ as units of distance and mass, respectively, and work at an
inverse temperature $\beta=14.7/W$ unless stated otherwise; this temperature is
low enough to identify the ordering tendencies in the model. 

\subsection{Methods}
We use DQMC to solve the 1D three-orbital HH model. The general details of the
method can be found in the Refs. [\onlinecite{BlankenbeclerPRD, WhitePRB,
ChangPMB}], while the aspects that are unique to the Holstein phonons can be
found in the Refs. [\onlinecite{JohnstonPRB2013}] and
[\onlinecite{ScalettarPRB1989}]. Throughout this work, we use a one-dimensional
chain with a chain size $N=16$ and imaginary time discretization of
$\Delta\tau=0.245/W$, unless otherwise stated. In all of our simulations, we
have not observed significant $\Delta\tau$ errors introduced by this choice. 

DQMC calculates the electron Green's function $G(k,\tau)$ defined in the imaginary time $\tau$ axis. 
In Sec. III. B we will examine the spectral properties of our model, which
requires an analytic continuation to the real frequency axis. Here, we used the maximum entropy method (MEM) [\onlinecite{JarrellPR,FuchsPRE}].

\section{Results}\label{Sec:Results}   
In this study, we will focus on the four phases resulting from the competition between the onsite Coulomb repulsion and the {\eph} interaction. The four phases are metal, charge-density-wave phase, orbital-selective Mott phase, and orbital-correlated state, as shown in Fig. \ref{Fig:Phase_sketch}. They are identified by examining the evolution of the single and double occupations on each orbital, the spectral weight, and the charge-density-wave susceptibility. We will discuss the particular variation of these quantities as a function of parameters in the following sections. 

\subsection{Weak Electron-Phonon Coupling}
\begin{figure}
\includegraphics[width=0.9\columnwidth]{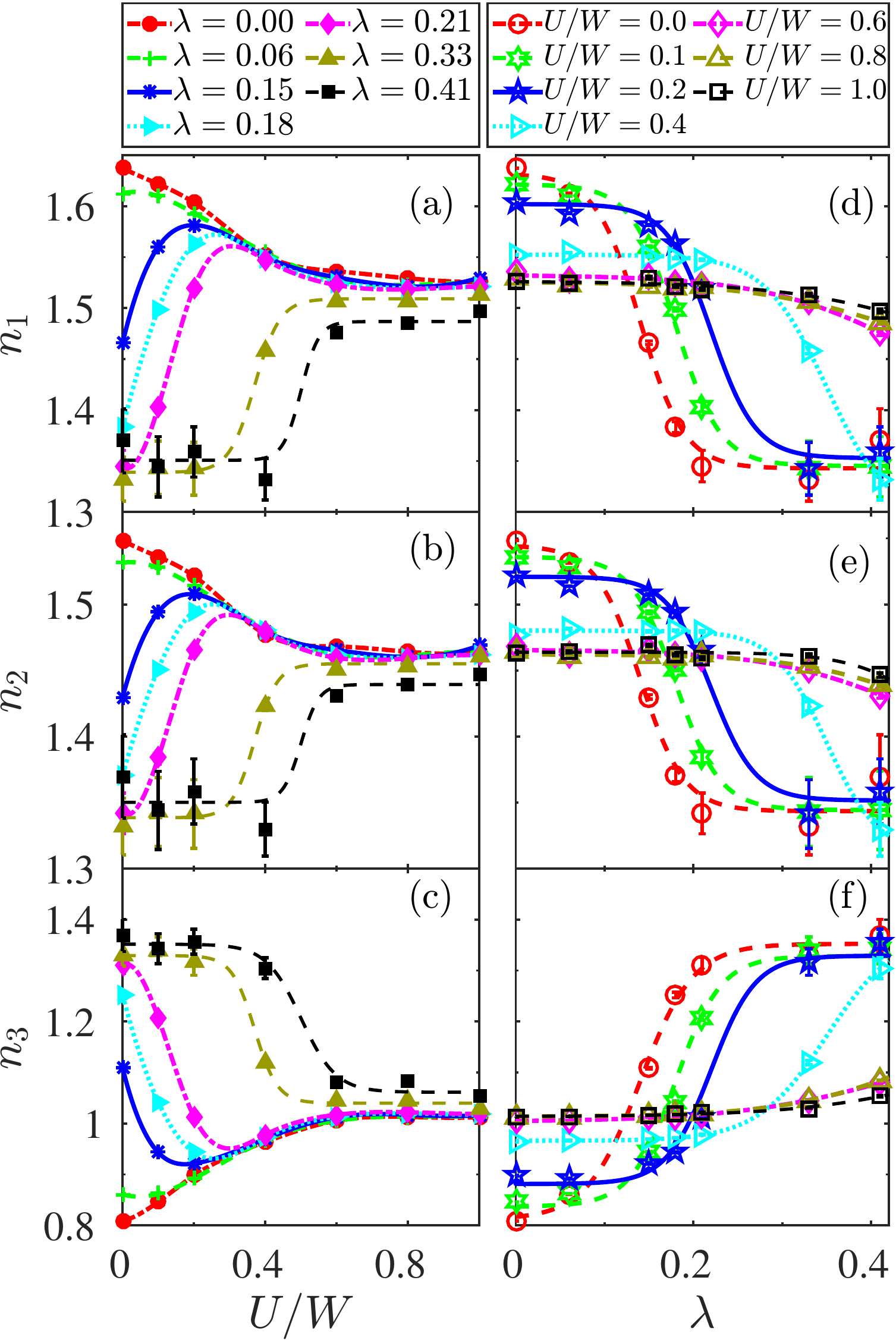}
 \caption{\label{Fig:fig1} (color online) The variation of electronic densities
on the three orbitals as a function of the Hubbard $U$ and the {\eph} coupling
strength $\lambda$. Panels (a) - (c) show the variation of electronic densities as a
function of the Hubbard $U$ on orbitals $\gamma = 1$, 2, and 3, respectively. 
Similarly, panels (d) - (f) show the change of electronic densities as a function of $\lambda$ on the same three orbitals. In each panel, error bars smaller than the marker size have been suppressed for clarity, and a smoothing spline is used as a guide to the eye.}
\end{figure}

\begin{figure}
 \includegraphics[width=0.9\columnwidth]{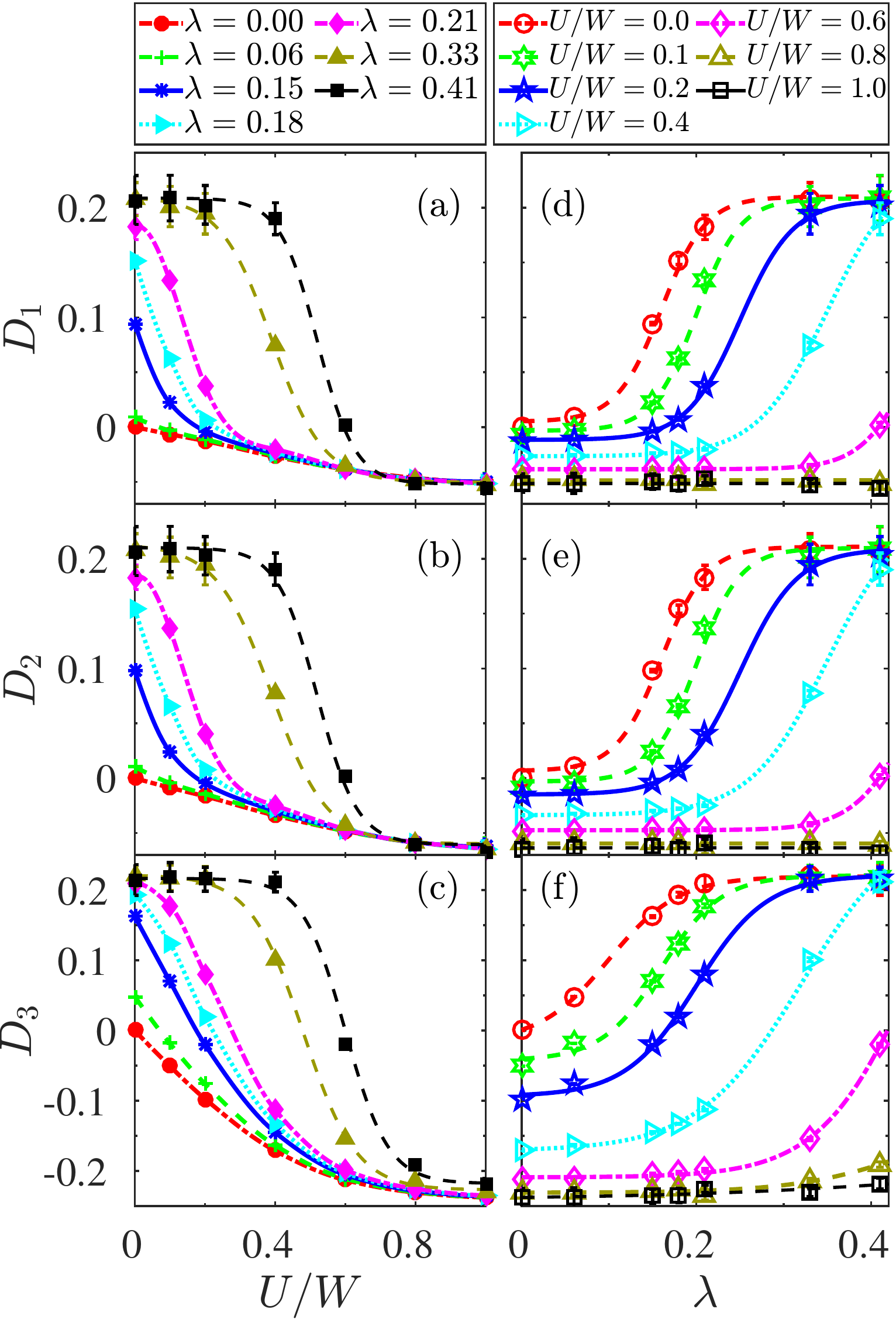}
 \caption{\label{Fig:fig2} (color online) The variation of double occupancies
on the three orbitals as a function of the Hubbard $U$ and the {\eph} coupling
strengths $\lambda$. Panel (a) - (c) shows the variation of double occupancies as a
function of the Hubbard $U$ on orbitals $\gamma = 1$, 2, and 3, respectively. 
Similarly, panels (d) -
(f) show the change of double occupancies as a function of $\lambda$ on the same three orbitals. 
 In each panel, error bars smaller than the marker size
have been suppressed for clarity, and a smoothing spline is used as a guide to the eye.
}
\end{figure}

\begin{figure}
 \includegraphics[width=0.7\columnwidth]{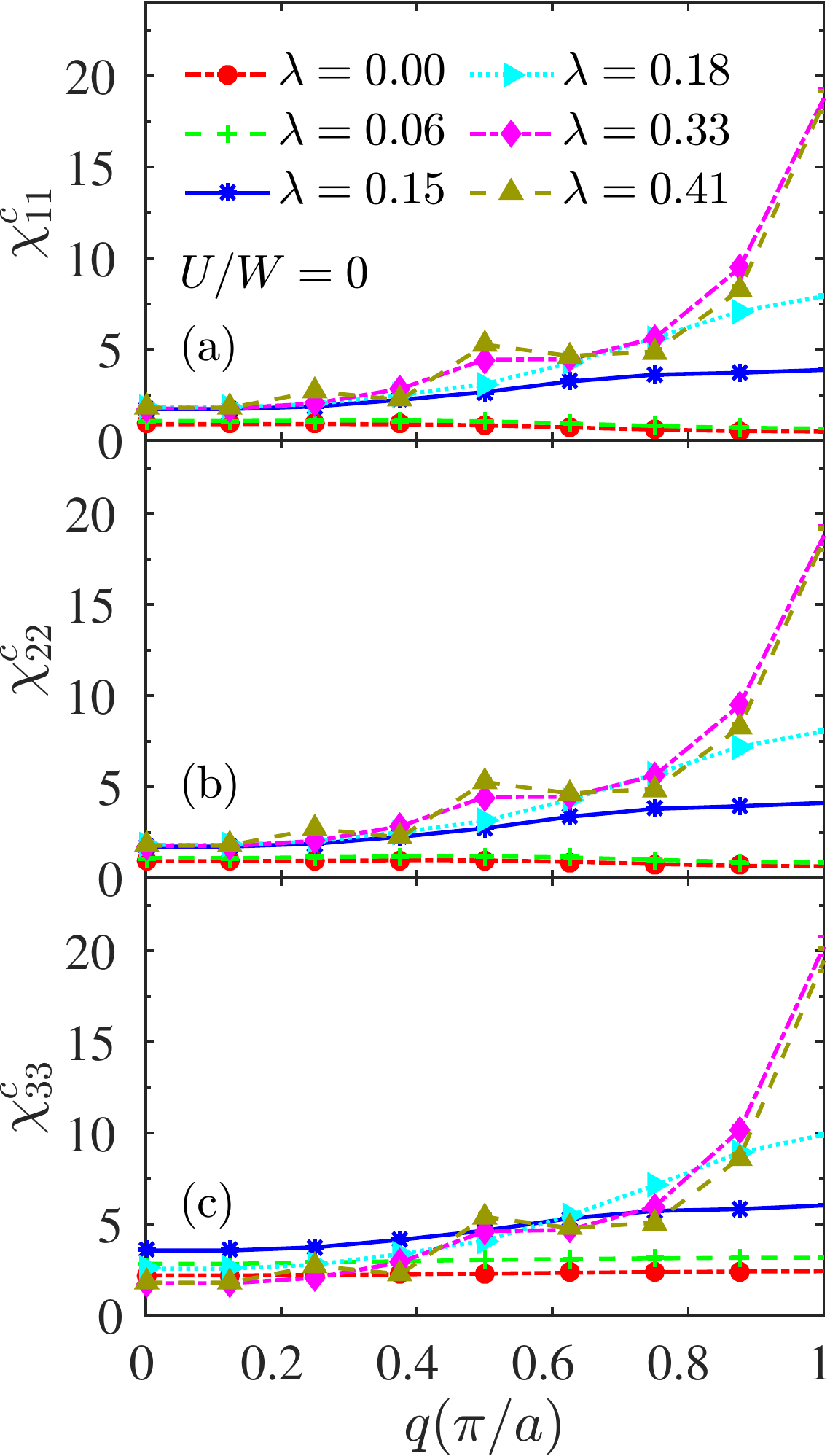}
 \caption{\label{Fig:chi_vs_lambda} (color online) Momentum dependence of the charge-density-correlation $\chi^c_{\gamma,\gamma}(q)$ for orbital 1 (a), orbital 2 (b), and orbital 3 (c) at different $\lambda$ values. The Hubbard $U/W=0$. In each panel, error bars smaller than the marker size have been suppressed for clarity.} 
\end{figure}

\begin{figure}
\includegraphics[width=0.65\columnwidth]{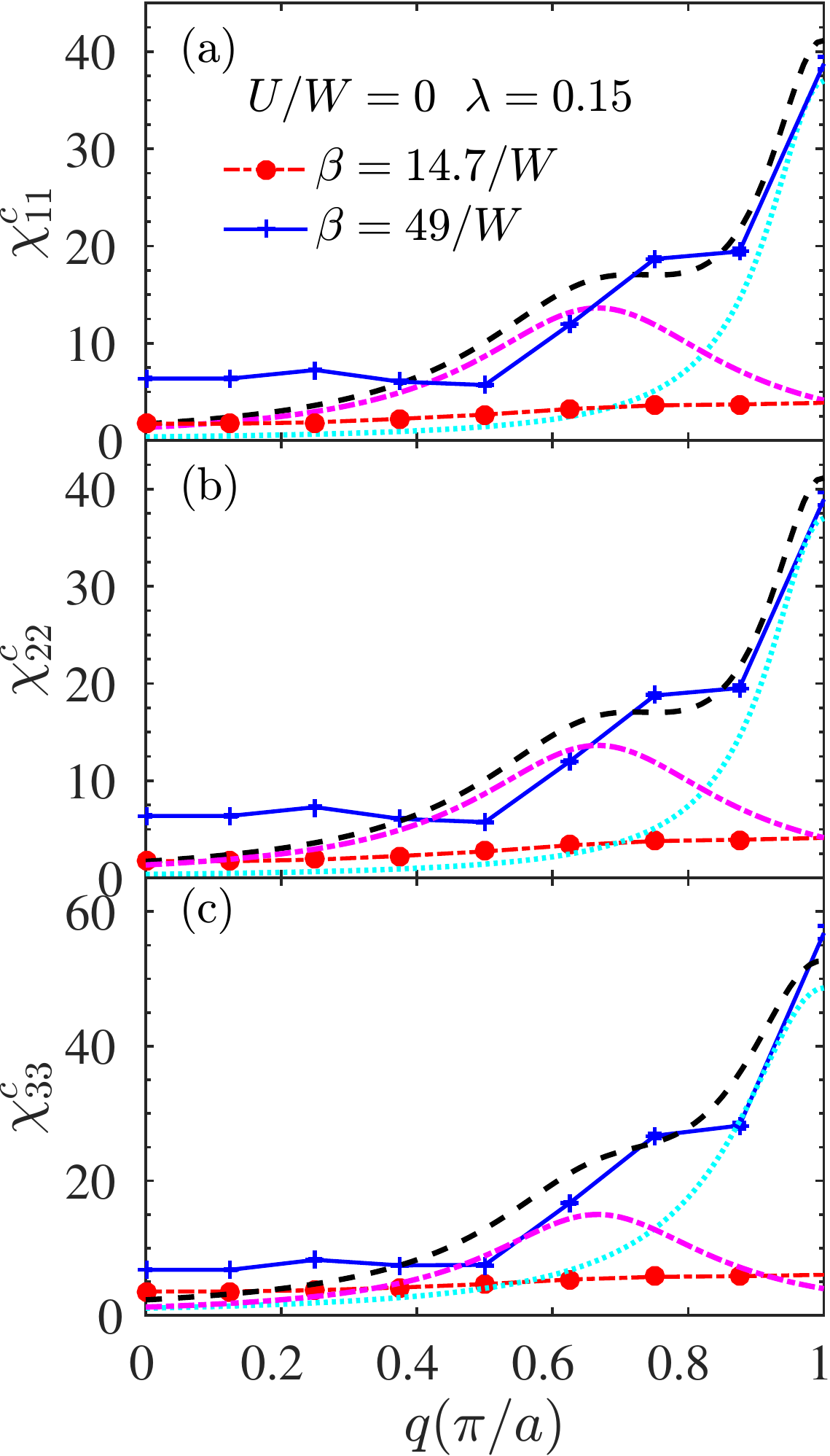}
 \caption{\label{Fig:fig3} (color online) Momentum dependence of the charge-density-correlation $\chi^c_{\gamma,\gamma}(q)$ for orbital 1 (a), orbital 2 (b), and orbital 3 (c) at different $\beta$ values. The Hubbard $U/W=0$ and $\lambda=0.15$. The black dashed lines are eye-guided lines for a combination of two Lorentzian functions with different peak positions. The two Lorentzian functions are shown with dotted and dash-dotted lines, respectively. In each panel, error bars smaller than the marker size have been suppressed for clarity.} 
\end{figure}

\begin{figure}
 \includegraphics[width=0.9\columnwidth]{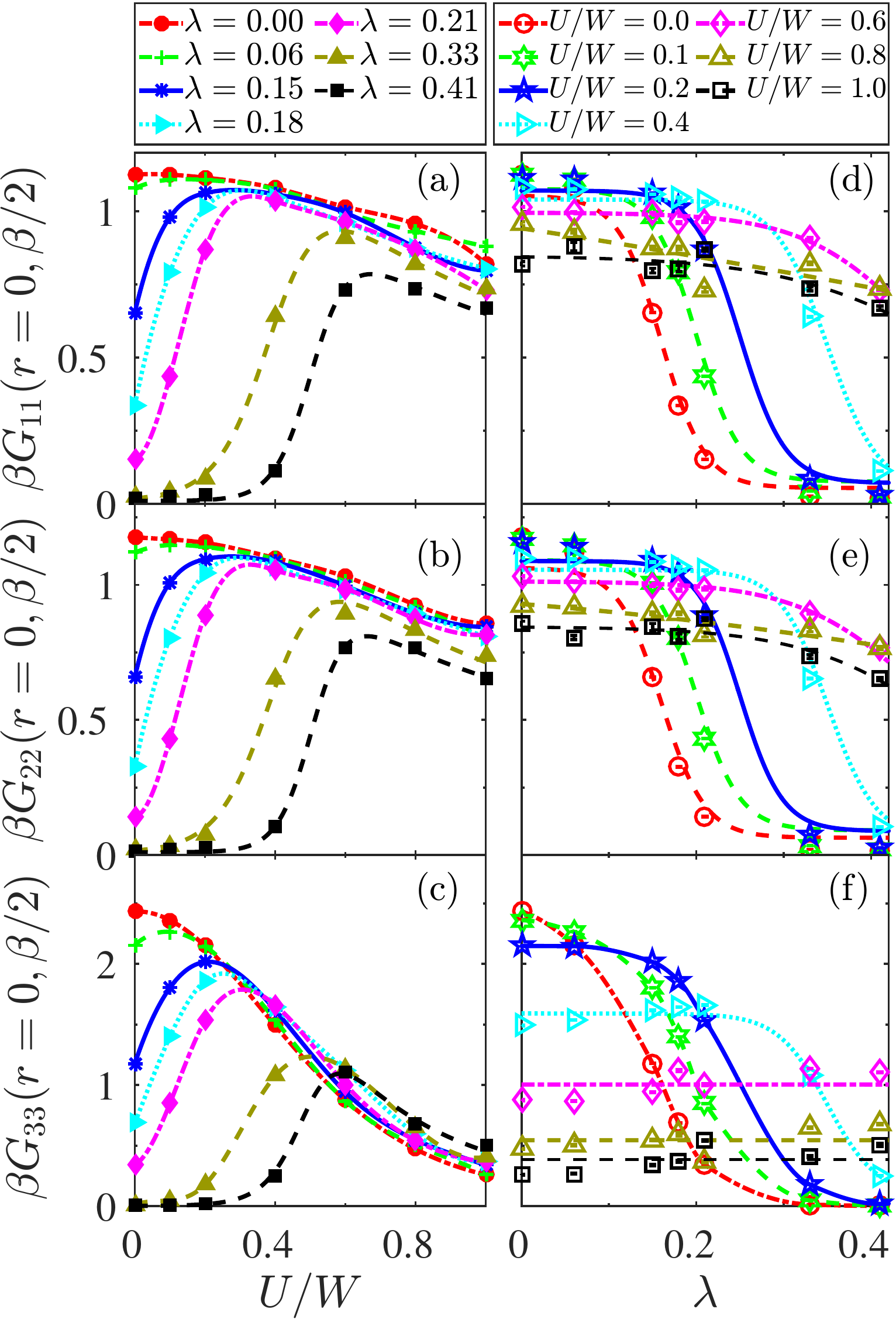}
 \caption{\label{Fig:fig4} (color online) The variation of the spectral weights for three orbitals as a function of the Hubbard $U$ and the {\eph} coupling strength $\lambda$. (a) - (c) show the variation of spectral weights as a function of the Hubbard $U$ for three orbitals, respectively. (d) - (f) show the change of spectral weights as a function of $\lambda$ for three orbitals, respectively. In each panel, error bars smaller than the marker size have been suppressed for clarity, and a smoothing spline is used as a guideline to the eye.}
\end{figure}

\begin{figure*}[t]
 \includegraphics[width=\textwidth]{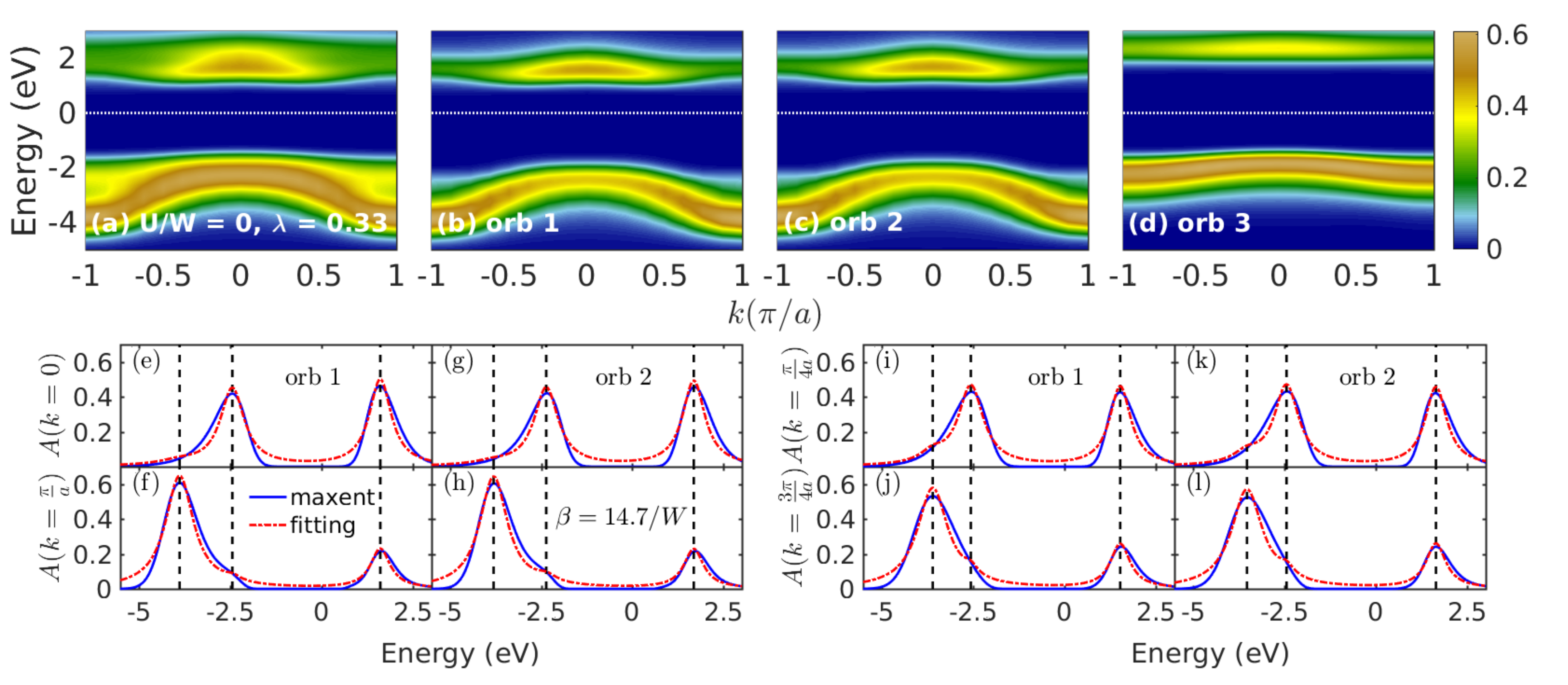}
 \caption{\label{Fig:fig5} (color online) Spectral functions for $U=0$ and $\lambda=0.33$. (b),(c), and (d) are the orbital 1, 2, and 3 parts of the spectral function in (a), respectively. (e), (f), (i), and (j) are spectral functions of the orbital 1 at momentum $k=0$, $k=\frac{\pi}{a}$, $k=\frac{\pi}{4a}$, and $\frac{3\pi}{4a}$, respectively. Similarly, (g), (h), (k), and (l) are spectral functions of the orbital 2 at those four momenta. The black dashed lines show three peaks positions in the maximum entropy results. The red dotted lines are Lorentzian fitting results. }
\end{figure*}

Previous studies of this model
without phonons showed that an OSMP forms for 
$U$ values in the range $0.6<U/W<2$ for our choice of $J$. In this state, 
orbital three becomes insulating while the remaining orbitals host itinerant
electrons [\onlinecite{RinconPRL,LiuPRE}]. The onset of this phase is signaled by the fact that the filling on orbital three $\langle \hat{n}_3 \rangle =1$. 
For $U/W > 2$, orbitals one and two retain a noninteger filling but are driven into an
insulating state by the onset of short-range orbital ordering
[\onlinecite{ShaozhiPRB2016}]. 
To avoid this complication, we restrict ourselves to the $U/W<1$ region. 

We now examine the impact of the {\eph} interaction on the OSMP. 
Figure \ref{Fig:fig1} plots the electronic occupations of the three orbitals for
different values of $U/W$ and $\lambda$. For $\lambda=0$, Fig. \ref{Fig:fig1}(c)
shows that $\langle \hat{n}_3\rangle$ converges to 1 as $U/W$ increases, 
implying that a Mott gap is formed on this orbital for $U/W \ge 0.4$.  
At the same time, $\langle \hat{n}_{1}\rangle$ and $\langle \hat{n}_2 \rangle$ 
maintain noninteger values, implying that these orbitals remain 
itinerant [see Fig. \ref{Fig:fig1}(a) and \ref{Fig:fig1}(b)]. 
These results are consistent with previous studies [\onlinecite{RinconPRL, LiuPRE}]. 
When we include the {\eph} coupling, the orbital occupations are modified significantly.  
For example, Figs. \ref{Fig:fig1}(d) - \ref{Fig:fig1}(f) show that the {\eph} coupling 
tends to make electronic occupations on all three orbitals uniform when $U/W<0.4$, 
with the average filling on each orbital approaching $\langle \hat{n}_\gamma\rangle = \frac{4}{3}$ 
when $\lambda$ is large. 
This value of the occupation on each orbital is consistent with a charge-ordered state where two sites are fully occupied, and one site is empty, which is shown in the CDW region of Fig. \ref{Fig:Phase_sketch}. This kind of charge order arises from the attractive interaction mediated by the {\eph} interaction [\onlinecite{PhilippePRB}]. 
For the fully occupied site, the attractive interaction can be mapped into a negative effective $U$ for all three orbitals at this site in the large $\Omega$ limit. For an empty site, the effective interaction is not modified by the e-ph coupling. Therefore, the effective Hubbard interaction is not uniform in the real space if there is a density modulation. When the e-ph coupling is strong, this nonuniform attractive interaction can produce a charge-ordered state but with a uniform average occupation on each orbital.
The transition from a CDW phase to an OSMP can be seen in
Fig. \ref{Fig:fig1}(c), where $\langle \hat{n}_3 \rangle$ decreases 
from $\frac{4}{3}$ to 1 at $\lambda=0.33$. Increasing the {\eph} coupling pushes this transition to larger values of $U/W$; for example, for $\lambda = 0.41$ it occurs at $U/W \sim 0.5$. 

The competition between the CDW and OSMP tendencies is also manifest 
in the behavior of the orbital's double occupation 
$D_{\gamma}=\langle n_{\gamma,\uparrow}n_{\gamma,\downarrow}\rangle-\langle
n_{\gamma,\uparrow}\rangle \langle n_{\gamma,\downarrow}\rangle$, 
as summarized in Fig. \ref{Fig:fig2}. When the phonon mediated effective attraction overcomes the Coulomb repulsion 
we expect $D_{\gamma}>0$; otherwise, $D_{\gamma}<0$. 
Figures \ref{Fig:fig2}(a)-\ref{Fig:fig2}(c) 
present $D_{\gamma}$ as a function of 
$U/W$ for fixed values of $\lambda$, where we find that $D_{\gamma}$ decreases as
$U$ is increased, and $D_3$ converges to
$-\frac{1}{4}$ in the limit of a strong Hubbard interaction, consistent with a Mott
insulating state where double occupation is suppressed. Figures \ref{Fig:fig2}(d)-\ref{Fig:fig2}(f)
alternatively plot the data as a function of $\lambda$ for fixed values of $U/W$. 
Here, we find that for $U/W<0.4$, $D_{\gamma}$
increases as $\lambda$ increases and converges to $\frac{2}{9}$ on each orbital. 
This value is consistent with the double occupations expected for the CDW phase shown in Fig. \ref{Fig:Phase_sketch}.  

The electronic density and double occupations provide indirect evidence of the
CDW phase. 
To obtain more direct evidence of a CDW order, we calculated the charge susceptibility 
\begin{eqnarray}
\chi^c_{\gamma,\gamma^\prime}(q)=\frac{1}{N} \int_0^{\beta} d \tau \langle \hat{n}_{q,\gamma}(\tau)\hat{n}_{q,\gamma^\prime}(0) \rangle,
\end{eqnarray}
where $q$ is the momentum, $\tau$ is the imaginary time, 
$\hat{n}_{q,\gamma}=\sum_{i,\sigma}e^{i q R_i}\hat{n}_{i,\gamma,\sigma}$, 
and $R_i$ is a lattice vector. 

Figure \ref{Fig:chi_vs_lambda} shows the momentum dependence of the three intraorbital 
charge susceptibilities for $U/W=0$ and 
different {\eph} coupling strengths. At weak coupling ({\it i.e.}  
$\lambda=0.0$ and $\lambda=0.06$), $\chi^c_{\gamma,\gamma}(q)$ is small,  
with no clear peak at any momenta. This observation implies that a finite 
value of $\lambda$ is needed for charge correlations 
to develop at this temperature, and is 
consistent with the one-dimensional Holstein model [\onlinecite{HardikarPRB2007}]. 
As the value of $\lambda$ is increased, a clear peak structure 
forms in $\chi^c_{\gamma\gamma}(q)$. For instance, already at 
$\lambda = 0.18$ we find a peak centered at $q=\pi/a$ for all three orbitals, 
indicating the formation of a two-sublattice charge correlation at $\beta = 14.7/W$.  
Upon further cooling of the system, we find that additional charge correlations 
develop at a second $q$ point. For example, 
Fig. \ref{Fig:fig3} compares $\chi_{\gamma,\gamma}^c(q)$ at 
$\beta=14.7/W$ and $\beta=49/W$ for the $\lambda = 0.15$ and $U/W = 0$ case. 
For temperature $\beta=14.7/W$, 
$\chi^c_{\gamma,\gamma}(q)$ has a single peak at $q=\pi/a$; however, 
as the temperature is decreased to $\beta=49/W$, $\chi^c_{\gamma,\gamma}(q)$ increases
and a second peak forms at $q \sim 2\pi/3 - 3\pi/4$, which is evident as a shoulder 
in $\chi^c_{\gamma,\gamma}(q)$. To better recognize these two peaks, we
plot as a guide-to-the-eye the sum of two Lorentzian functions centered 
at $q=2\pi/3a$ (dash-dot line) and $q=\pi/a$ (dotted line). 

%This two peak structure in $\chi^c_{\gamma,\gamma}(q)$ likely reflects different ordering tendencies established by the nesting conditions across the three bands crossing the Fermi level in the non-interacting model. 
The two peak structures in $\chi^c_{\gamma,\gamma}(q)$ likely reflect different ordering tendencies. 
The charge configurations sketched in Fig. \ref{Fig:Phase_sketch} are 
consistent with $q_1\approx2\pi/3a$ and $q_2=\pi/a$ orderings. We, therefore, propose 
that the CDW state is characterized by a superposition of 
$|\dots 6 6 0 6 6 0 \dots \rangle$ and $|\dots 6 0 6 6 6 0 \dots \rangle$ 
configurations along the chain, where the number indicates the number of carriers on each site. 
These charge configurations are consistent with the values of the orbitally-resolved 
single and double occupancies discussed previously. 
In fact, these two peaks reflect two different values of $2k_\mathrm{F}$ that appear in this multiorbital model; 
The Fermi momentum for orbitals 1 and 2 are $\sim 0.33\pi/a$, while the Fermi momentum for orbital 3 is $\sim0.5\pi/a$. 
Thus, these two peak values correspond to $q = 2k_\mathrm{F}$ in the weak coupling limit, where the CDW tendencies are driven primarily by nesting conditions.  
When we increase the {\eph} coupling further, the $k_\mathrm{F}$ for orbitals 1 and 2 increases to $0.5\pi/a$. 
Therefore, we expect that only one peak will be observed in the charge-density-wave susceptibility in the limit 
of strong {\eph} coupling.

We now turn to the spectral weight of the three orbitals in the vicinity of
the Fermi level $E_\mathrm{F}$ to assess whether the various phases we observe are insulating or not. 
The spectral weight can be estimated directly from the imaginary time Green's
function using the relationship [\onlinecite{TrivediPRL}]
\begin{eqnarray}
\beta G_\gamma(r=0,\beta/2)=\beta\sum_k \int d\omega \ \mathrm{sech}(\beta\omega/2)A_\gamma(k, \omega),
\end{eqnarray}
where $A_\gamma(k, \omega)$ is the orbitally-resolved spectral function. 
At low temperature, the function $\beta
\mathrm{sech}(\beta\omega/2)$ is sharply peaked around $\omega=E_\mathrm{F}=0$ and thus 
provides a measure of the spectral weight integrated over a window of a few $\beta^{-1}$ of $E_\mathrm{F}$.

Figure \ref{Fig:fig4} plots $\beta G_\gamma(r=0,\beta/2)$ for the 
three orbitals for different values of $U/W$ and $\lambda$. 
In Figs. \ref{Fig:fig4}(a)-\ref{Fig:fig4}(c), the spectral weight of each 
orbital is plotted as a function of $U$ for fixed $\lambda$. In the absence of
the {\eph} interaction ($\lambda = 0$), the spectral weight of all three
orbitals decreases as the Hubbard $U$ increases. However, the spectral
weight on orbital $\gamma = 3$ decreases more rapidly than the other two
orbitals, 
consistent with the formation of an OSMP [\onlinecite{ShaozhiPRB2016}].
[The small but nonzero value of $\beta G_{3}(r=0,\beta/2)$ is due to
the elevated temperature of the simulation.] As the value of $\lambda$
increases, we begin to see the loss of spectral weight in all three orbitals
when $U/W$ is small. 

Fig. \ref{Fig:fig4}(d)-\ref{Fig:fig4}(f) plot  
$\beta G_\gamma(r=0,\beta/2)$ as a function of $\lambda$ for fixed $U/W$. 
For $U/W = 0$, the spectral weight of all three orbitals is suppressed as the {\eph} coupling is increased. We further observe a sudden decrease in the spectral weight of all three orbitals for $\lambda \ge 0.15$, 
where a prominent peak in $\chi_{\gamma,\gamma}^c(k)$ is observed. 
Thus, for $\lambda \ge 0.15$ and 
$U/W = 0$, the system is an insulating state driven by CDW correlations.  
The behavior of the spectral weight is qualitatively similar for $U/W < 0.2$, however, the transition to the CDW phase occurs at larger values of $\lambda$ as $U/W$ increases. 
Based on these results, we conclude that the CDW phase is insulating. 
 
Unlike the OSMP, we do not find any orbital-selective behavior associated with the formation of the CDW phase; the rate at which the spectral weight approaches zero appears to be the same for all three orbitals at this temperature. This result is in contrast to the degenerate two-orbital case with inequivalent bandwidths [\onlinecite{Shaozhiarxiv}], where orbital selective CDW behavior was found. 
This difference could be attributed to changes in the total bandwidth, dimensionality, or model. 
(For example, the current model has inequivalent bandwidths and crystal field splittings while the former only had 
inequivalent bandwidths.) Further studies will be needed to better understand the differences between these two cases. 
 
We now return to the competition between the OSMP and CDW phases. 
Fig. \ref{Fig:fig4}(a)-(c) reveals that the spectral weight 
decreases as the strength of the {\eph} is increased when 
$U/W$ is small. For a fixed value of $\lambda\ne 0$, $\beta G_\gamma(r=0,\beta/2)$ 
initially increases with $U/W$, before reaching a maximum value and 
decreases along the $U$ axis. This behavior reflects the competition between 
the {\eph} and the {\ee} interactions [\onlinecite{NowadnickPRL,NowadnickPRB}]. At small $U/W$ the CDW correlations dominate, 
for large $U/W$ the OSMP correlations dominate, and for intermediate values of $\lambda$ a metallic phase is realized. 

\subsection{Spectral Properties of the CDW Phase}
\begin{figure*}[t]
 \includegraphics[width=\textwidth]{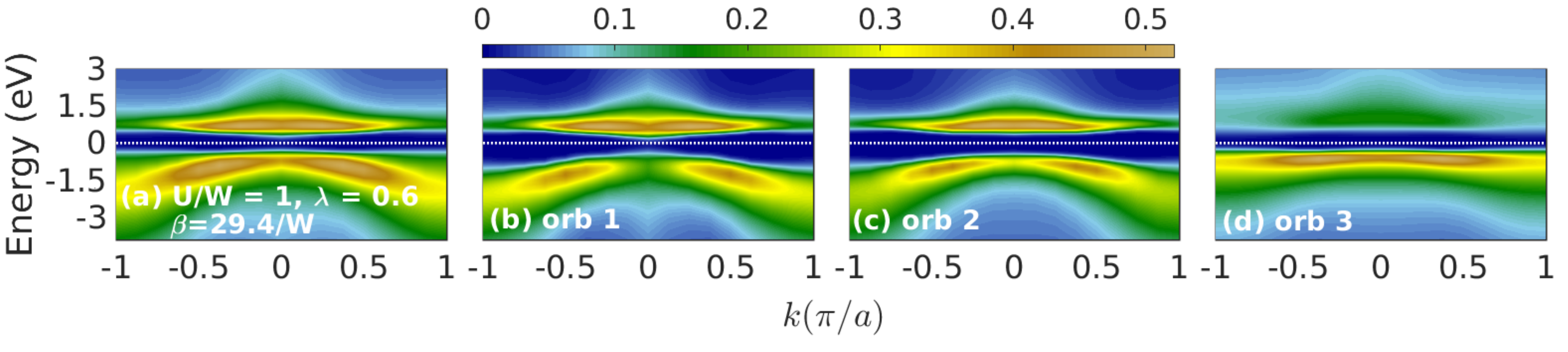}
 \caption{\label{Fig:fig6} (color online) Spectral functions for $U/W=1$ and $\lambda=0.6$. (b), (c), and (d) are the orbital 1, 2, and 3 parts of the spectral function in (a). The inverse temperature is $\beta=29.4/W$. The white dot line is the Fermi surface.}
\end{figure*}

The spectral function of the OSMP was studied in detail in Ref.
[\onlinecite{ShaozhiPRB2016}] in the absence of {\eph} interactions. We will, therefore, focus on the spectral function of the CDW phase. 
Figure \ref{Fig:fig5} shows the spectral function for $U=0$ and $\lambda=0.33$
and its orbitally-resolved components. The system is insulating, 
with a large CDW gap and broadened spectral features, consistent with our spectral weight analysis. The upper bands of three orbitals have dispersions with a 
clear folded shape, while the lower bands
of orbitals 1 and 2 have a more cosine-like dispersion. 
This cosine-like shape arises from the combination of an incoherent peak and an additional 
peak arising from thermally activated transitions to states with additional phonons excited
[\onlinecite{HohenadlerPRB}]. To better recognize these two peaks, Figs.
\ref{Fig:fig5}(e) -  \ref{Fig:fig5}(l) plot the spectral functions at fixed 
momentum $k=0$, $\pi/a$, $\pi/4a$, and $3\pi/4a$ for orbitals 1 and  2. 
The red dashed curve 
are Lorentzian fits of the data allowing for an incoherent peak above and below the
Fermi level and an additional thermally excited peak below the Fermi level. The fitting
results are consistent with the Maxent results. We find that the thermally
excited state is located around $E=-2.5$ eV and is momentum independent, 
consistent with previous results for the one-dimensional single-band spinless
Holstein model [\onlinecite{HohenadlerPRB}]. The folded band is observed at
$k=0$ and $k=\pi/a$ and at $k=\pi/4a$ and $k=3\pi/4a$, respectively. The
intensity of the incoherent peak below the Fermi level in the folded band is
much weaker than that of the thermally excited peak, leading to a cosine shape
observed in the upper panels of Fig. \ref{Fig:fig5}.

\subsection{Strong electron-phonon coupling}
\begin{figure}
 \includegraphics[width=\columnwidth]{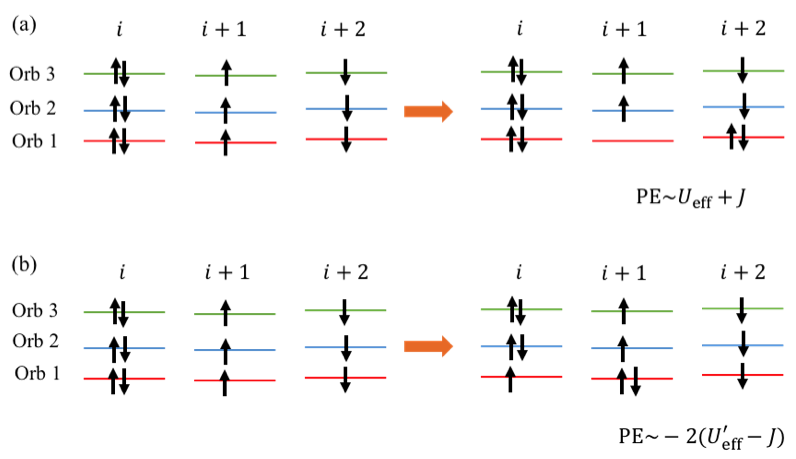}
 \caption{\label{Fig:fig8} (color online) Cartoon sketch of the relevant charge fluctuation processes leading to the orbital ordered insulating state when $U/W=1$ and $\lambda=0.6$.}
\end{figure}

Our previous single-site DMFT study of the two-orbital HH model at a half
filling observed a direct transition between the OSMP and CDW phase in the
strong {\eph} coupling limit [\onlinecite{Shaozhiarxiv}], with no intervening
metallic phase. In contrast, for the current model, we find evidence for an
orbitally correlated insulating state
located between the CDW phase and OSMP at strong couplings. 
Figure \ref{Fig:fig6} shows the spectral function [Fig. \ref{Fig:fig6}(a)]
and its three orbitally-resolved components [Figs. \ref{Fig:fig6}(b) - \ref{Fig:fig6}(d)] at 
$U/W=1$, $\lambda=0.6$, and $\beta=29.4/W$, where 
a gap is clearly observed.  
The spectral function is similar to the CDW case shown in Fig. \ref{Fig:fig6}, however, 
the origin of the gap is not CDW correlations since 
$\chi^c_{\gamma,\gamma^\prime}(q)$ (not shown)  
is small in this state. Also, the double occupation
$D_\gamma\approx-\frac{1}{9}<0$ (see Fig. 2), indicating the Coulomb interaction is the dominant 
interaction in this phase.

The nature of this OC phase is sketched on the left side of Fig. \ref{Fig:fig8}. 
It consists of one site where all three orbitals are fully occupied and two
neighboring sites that are half filled and in a high-spin state. 
This electronic configuration is consistent
with the observed orbital occupations and the value of the double
occupation $D_\gamma=-\frac{1}{9}$. 

In the HH model, the intra- and interorbital Coulomb interactions $U_\mathrm{eff}$ and
$U^\prime_\mathrm{eff}$ are renormalized by the {\eph} interaction, and the
ground state can change based on the value of these effective Coulomb interactions. Figures  
\ref{Fig:fig8}(a) and \ref{Fig:fig8}(b) show two types of charge fluctuations 
that are possible within the proposed OC state. 
The potential energy cost of these fluctuations are
$PE\sim U_\mathrm{eff}+J$ and $PE\sim-2(U^\prime_\mathrm{eff}-J)$,
respectively. To estimate the magnitude of these energies, we performed an exact diagonalization calculation in the atomic limit and compared the ground state energies of the shown atomic configurations. For $U/W=1$ and $\lambda=0.6$, 
we find that these two potential energies are 1.305 eV
and 0.815 eV, respectively. When the orbital hybridization is introduced, the total potential energy cost is compensated for by a kinetic energy gain of $KE\approx t_{11}=0.5$~eV. 
However, the ratio $\frac{PE}{KE}>1$ in both cases, suggesting that charge fluctuations are suppressed, and the system will be insulating.  
The conditions for forming the OC insulating state are then 
$U^\prime_\mathrm{eff}-J<0$ and $U_\mathrm{eff}+J>0$, which in turn requires that the {\eph}
coupling strength is not too strong; otherwise, the CDW phase is formed.  
(Note that a larger Hund's coupling favors satisfying these two conditions.) 
In the OC insulating state, the fully occupied site and two half occupied
sites can be arranged randomly in a long chain as the energy cost will not change. Therefore, short-range orbital correlations would be sufficient to produce insulating behavior at finite temperature.  

\begin{figure}
 \includegraphics[width=0.9\columnwidth]{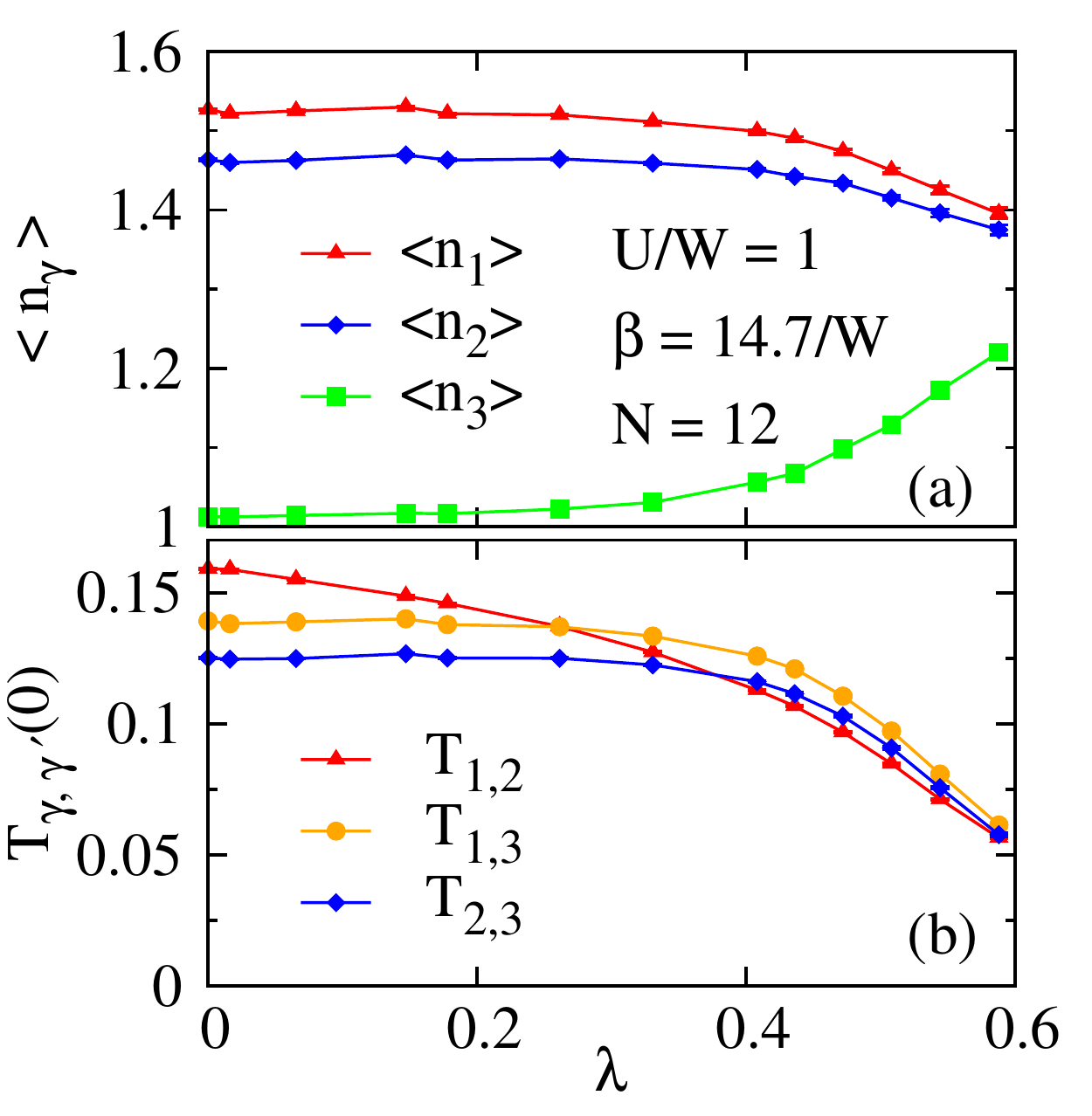}
 \caption{\label{Fig:fig9} (color online) (a) The electronic occupations $\langle\hat{n}_\gamma\rangle$ of orbital $\gamma$ as a function of the {\eph} coupling strength $\lambda$. (b) The onsite orbital correlations $T_{\gamma\gamma^\prime}$ between orbital $\gamma$ and $\gamma^\prime$ as a function of $\lambda$. These results were obtained on an $N=12$ site chain. In each panel, error bars smaller than the marker size have been suppressed for clarity. }
\end{figure}

We can confirm the presence of OC by examining the equal time 
orbital correlation function 
\begin{eqnarray}
T_{\gamma\gamma'}(d)=\frac{1}{N}\sum_i\langle (\hat{n}_{i+d,\gamma}-\hat{n}_{i+d,\gamma^\prime})(\hat{n}_{i,\gamma}-\hat{n}_{i,\gamma^\prime})\rangle.
\end{eqnarray}
The charge configuration expected for the OC shown in Fig. \ref{Fig:fig8} would produce an 
orbital correlation function $T_{\gamma\gamma'}(d)=0$. 
Figure \ref{Fig:fig9}(b) shows the variation of the onsite orbital
correlations $T_{\gamma,\gamma'}(0)$ in the phase transition from the OSMP to
the OC state. In general, the correlation function for $d=0$ is larger than that for $d\ne 0$. 
Thus, the onsite correlation is a good indicator for when $T_{\gamma,\gamma'}(d) \rightarrow 0$ and can 
be used to trace the forming of the orbital correlations. 
In the OSMP region, $T_{\gamma,\gamma^\prime}(0)$ is greater than 0.1. For
example, at $\lambda=0$, $T_{1,2}(0)$, $T_{1,3}(0)$, and $T_{2,3}(0)$ are 0.16,
0.14, and 0.125, respectively. $T_{\gamma,\gamma^\prime}(0)$ decreases slowly
initial as the {\eph} coupling strength is enhanced. Conversely, near the phase
transition, $T_{\gamma,\gamma^\prime}(0)$ decreases very quickly; at
$\lambda=0.6$, $T_{1,2}(0)$, $T_{1,3}(0)$, and $T_{2,3}(0)$ are 0.056, 0.061,
and 0.057, respectively. The nonzero value at $\lambda=0.6$ is likely due to the
elevated temperature. We find $T_{1,2}(0)$, $T_{1,3}(0)$, and $T_{2,3}(0)$ are
decreased to 0.038, 0.03, and 0.029, respectively, as the inverse temperature
is decreased to $29.4/W$. 
We expect that the correlation function tends towards zero as $T \rightarrow 0$, and 
a sharp phase transition from the OSMP to the OC state would occur. 

We traced the phase transition from the OSMP to the OC in Fig. \ref{Fig:fig9}(a), which shows the variation of orbital occupations in the phase transition from the
OSMP to the OC phase at $U/W=1$ and $\beta=14.7/W$. Here, the chain size is $N=12$. The critical {\eph} coupling value $\lambda_c$ of the phase
transition from the Mott phase to the OC state is about 0.43, where $n_3>1$. As
$U/W=1$ and $\lambda=0.6$, orbital occupations for three orbitals are 1.4,
1.38, and 1.22 at $\beta=14.7/W$, respectively. Those occupations are changed
to 1.346, 1.344, 1.31 at $\beta=29.4/W$, implying the OC state supports the
same occupation on each orbital, consistent with the electron configuration
shown in Fig. \ref{Fig:fig8}.

\section{Discussion And Summary}

\begin{figure}
 \includegraphics[width=\columnwidth]{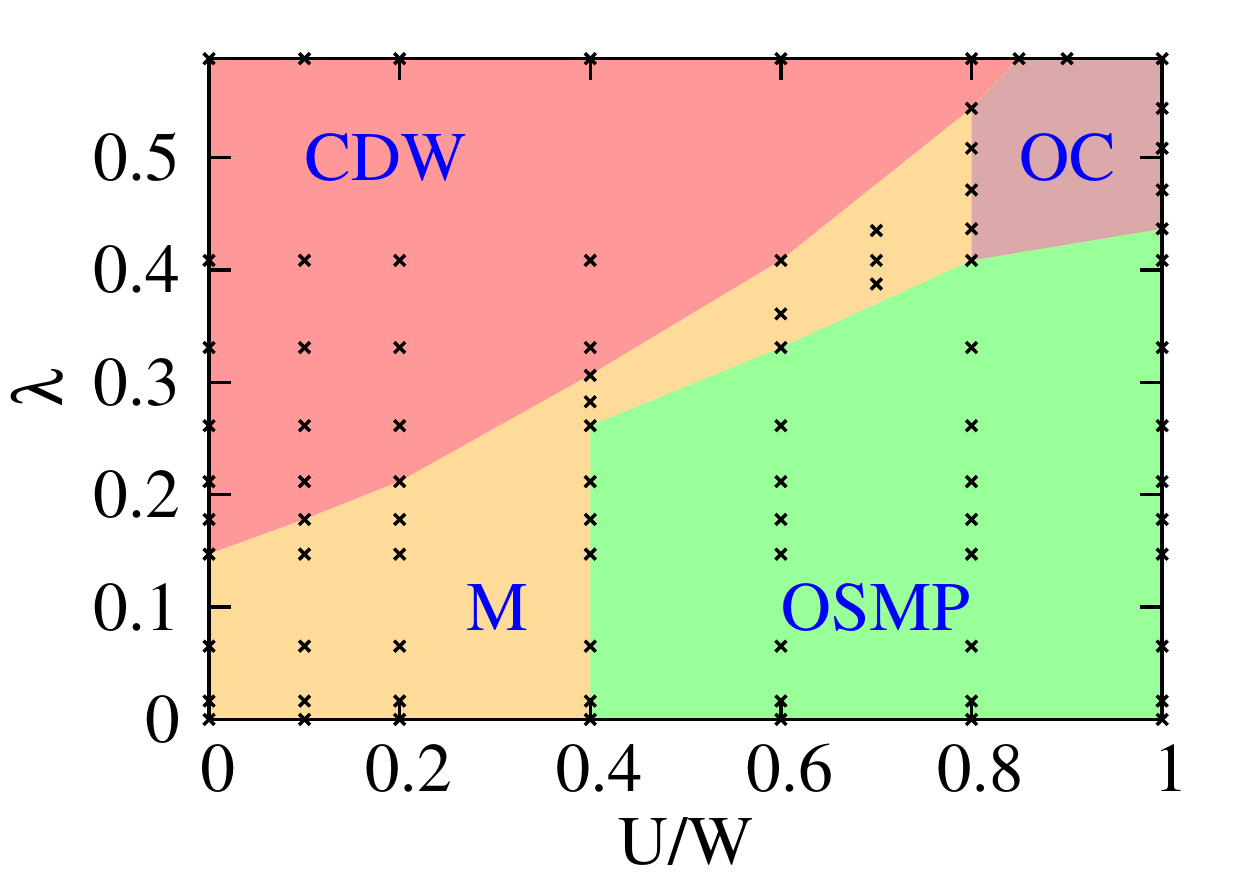}
 \caption{\label{Fig:fig10} (color online) The phase diagram of the three-orbital Hubbard-Holstein model for $\beta=14.7/W$ and $n=4$. The different phases are labeled as follows: metal (M), charge-density-wave order (CDW), orbital selective Mott phase (OSMP), and orbital correlation(OC).}
\end{figure}

We have performed a study of a three-orbital Hubbard-Holstein model on an
extended one-dimensional chain using non-perturbative DQMC. The phase diagram
of the one-dimensional model for $\beta=14.7/W$ and $n=4$, shown in Fig.
\ref{Fig:fig10},  
shares many similarities to the one found for an infinite dimensional 
degenerate two-orbital HH model with inequivalent bandwidths 
[\onlinecite{Shaozhiarxiv}], containing a metallic phase, a CDW phase, and an OSMP. The metallic phase is most prominent in small values of 
$U$ and $\lambda$ but penetrates into the region of intermediate interaction strengths separating the CDW and OSMP when $U/W \sim 2\lambda$. The critical {\eph} coupling 
needed for the CDW phase transition at $U=0$ in our model is $\lambda_c \sim 0.15$. 
The nonzero value of $\lambda_c$ is consistent with results for the single-band Holstein model [\onlinecite{HirschPRL,HardikarPRB2007}]. 

At strong couplings, we found evidence for an orbital correlation state in the phase diagram, which was not found in the previous DMFT study. We argue that 
this difference stems from the filling used in our model ($\langle \hat{n}\rangle = 4$ here versus 
$\langle \hat{n}\rangle = 2$ in Ref. [\onlinecite{Shaozhiarxiv}]) and the use
of an extended cluster here [\onlinecite{ShaozhiPRB2016}]. The OC state resides
between
the CDW phase and OSMP and tends to extend to large Hubbard $U$. 
This region of the phase diagram is the same one where the
OSMP disappears, and an anti-ferro-orbital correlation was found in the $\lambda=0$ case [\onlinecite{ShaozhiPRB2016}]. We expect that a
phase transition occurs between the anti-ferro-orbital order and the OC state
at $U/W=2$. These results show that
the phase diagram of multiorbital HH models can exhibit remarkably rich
physics as a function of interaction strengths, doping, and other parameters. 

In the past, the {\eph} interaction has been neglected when studying the
FeSCs. As mentioned in the introduction, this is largely motivated by early
{\it ab initio} calculation showing that the dimensionless interaction strength
$\lambda$ is only about 0.2 [\onlinecite{BoeriPRL2008}]. However, we find the
electronic degrees of freedom in our 1D multiorbital model can be influenced
significantly by small {\eph} couplings. Given that a similar result was obtained in the
case of a two-orbital Hubbard model in infinite dimensions using DMFT
[\onlinecite{Shaozhiarxiv}], these results imply that the {\eph} interaction cannot be
neglected {\it a priori} in multiorbital systems, and that this conclusion
holds independent of the dimension of the system. We hope that this study will
motivate further work in this interesting area. 

\section{Acknowledgements}
This work was supported by the Scientific Discovery through Advanced Computing
(SciDAC) program funded by U.S. Department of Energy, Office of Science,
Advanced Scientific Computing Research and Basic Energy Sciences, Division of
Materials Sciences and Engineering.
CPU time was provided by resources supported by the University of Tennessee and
Oak Ridge National Laboratory Joint Institute for Computational Sciences
(http://www.jics.utk.edu).

\end{document}